\documentclass{ws-mpla}
\usepackage[super]{cite}
\usepackage{graphicx}
\usepackage[colorlinks,linkcolor=blue,citecolor=red]{hyperref}
\usepackage{titlesec}
\usepackage{hyperref}
\usepackage{tabularx}
\usepackage{booktabs}
\newcommand{\be}{\begin{eqnarray}}
\newcommand{\ee}{\end{eqnarray}}

\def\be{\begin{equation}}
\def\ee{\end{equation}}
\def\beg{\begin{align}}
\def\eeg{\end{align}}
\def\bea{\begin{eqnarray}}
\def\eea{\end{eqnarray}}

\def\l{\left}
\def\r{\right}

\begin{document}

\markboth{Sobouti \& Sheikhahmadi}{Remove point-mass concept-remove  singularities from GR}

%%%%%%%%%%%%%%%%%%%%% Publisher's Area please ignore %%%%%%%%%%%%%%
\catchline{}{}{}{}{}
%%%%%%%%%%%%%%%%%%%%%%%%%%%%%%%%%%%%%%%%%%%%%%%%%%%%%%%%%%%%%%%%%%%

\title{Remove point-mass concept-remove  singularities from GR}

\author{Yousef Sobouti}

\address{Institute for Advanced Studies in Basic Sciences, 444 Prof. Yousef Sobouti Blvd., Zanjan 45137-66731, Iran\\
Fellow, Academy of Sciences of Iran\\
sobouti@iasbs.ac.ir}

\author{Haidar Sheikhahmadi}

\address{School of Astronomy, Institute for Research in Fundamental Sciences (IPM),  P. O. Box 19395-5531, Tehran, Iran\\
h.sh.ahmadi@gmail.com;h.sheikhahmadi@ipm.ir
}

\maketitle

%

%\tableofcontents
%\pub{Received (Day Month Year)}{Revised (Day Month Year)}
%
\begin{abstract}
 Singularities in Newton's gravitation,  in general relativity (GR), in Coulomb's law,  and elsewhere in classical physics, stem from two ill conceived assumptions:  a) there are point-like entities with finite masses, charges, etc., packed in zero volumes, and b)  the non-quantum assumption that these point-likes can be assigned precise coordinates and momenta. In the case of GR, we argue that the classical energy-momentum tensor in Einstein's field equation is that of a collection of point particles and is prone to singularity.
 In compliance with Heisenberg's uncertainty principle,  we suggest to replace each constituent of the gravitating matter with a suitable quantum mechanical equivalent, here a Klien-Gordon (KG) or a Yukawa-ameliorated version of it, YKG field. KG and  YKG  fields are spatially distributed entities. They do not end up in singular spacetime points nor predict singular blackholes.
 On the other hand, YKG waves reach infinity as
  $\frac{1}{r}e^{-(\kappa\pm i k)r}$.
  They create the Newtonian $r^{-2}$  term as well as a non-Newtonian  $r^{-1}$ force. The latter is capable of explaining the observed flat rotation curves of spiral galaxies,  and is interpretable as an alternative gravity, a dark matter scenario, etc.  There are ample observational data on flat rotation curves of spiral galaxies, coded in the Tully-Fisher relation, to support our propositions.\\
  \\
\noindent{\emph{ Dedicated to  International Year of Quantum Science and Technology 2025}.}

\ccode{PACS Nos.: $04.50.Kd$ Modified theories of gravity; $04.60.Bc$ Phenomenology of quantum gravity; $04.70.-s$ Physics of blackholes}
\vspace{0.3cm}
\end{abstract}

\newpage
\section{Introduction}\label{introduction}

In Newton's gravitation, $GM/r$, or in Coulomb's law, $Q/r$, one implicitly assumes there are physical entities of finite mass or charge packed in zero volumes. The assumption further implies that these point-likes can be assigned precise coordinates and momenta, and when approached infinitely closely, produce infinite gravitation or electric fields.
That Newton's and Coulomb's laws have played decisive roles in setting the physics of the 17th-19th Centuries on  axiomatized mathematical foundations is well acknowledged. That they are remarkably accurate to meet the everyday needs of the 21st century, from laboratory measurements to the solar system and galactic observations, is remarkable.
The notion of  point-like entities, however, is at odds with the foundations of quantum mechanics.  Heisenberg's uncertainty  principle  does not allow the assignment of precise coordinates and/or momenta to a physical entity. All one  can  hope for,  is to find the particle, or whatever it is, in a quantum cloud of probability in a phase space.

The point particle singularity is a common feature of all non-quantum physics, ranging from classical mechanics, classical electromagnetism, special and general relativity, to the physics of continuous media in its broadest sense. As long as the phase-space, ($\mathbf{x}\otimes\mathbf{p}$),  available to a dynamical system is spacious,   classical approximations are adequate.  Deviations,  however,  appear when a system is forced to evolve in tighter and tighter phase volumes.  Common examples are the cases of atomic and molecular spectroscopy,  where electrons are forced to stay bound to the nuclei of their host atoms or molecules with minimal energies. Another example from astronomical realms is the neutron star,  an aggregate of neutrons believed to be compressed into exceedingly small volumes by gravitational forces and cooled to about Fermi temperatures.  Then there follows the popular conclusion that,  if collapsed, the neutron star becomes a singular blackhole,  a spacetime singularity.

This paper builds upon our ongoing effort to address the limitations of the point-particle concept.
 In \cite{Sobouti:2023kkb}, one of us removes the Coulomb singularity of the Dirac electron by proposing a mutual action-reaction partnership between the Dirac wave function and the electric field of the electron itself. By so doing he comes up with a distributed charge and current for the spinning electron and the correct gyromagnetic ratio, however, without recourse to the QED formalism.
In the case of GR, we argue that the classical energy-momentum tensor in Einstein's field equation, no matter how one conceives it, is a collection of inherently singular point particles.
%This inevitably entails the non-quantum act of assigning precise coordinates and momenta to the particles.
In compliance  with the uncertainty principle, we suggest to replace (not the lumpsum of the gravitating matter but) each member in that collection of singulars  by a quantum mechanical equivalent each represented by a KG or by a  normalizable  YKG field, say.

Resorting to auxiliary fields in GR,  for different needs and purposes,  has a long history and rich literature. Pioneering work by Ruffini and Bonazzola \cite{Ruffini:1969qy} investigated systems of self-gravitating scalar bosons and spin-$1/2$ fermions. Based on numerical solutions,  they conclude that spacetime remains  non-singular. Buchdal \cite{Buchdahl:1959zz} and  Virbhadra \cite{Virbhadra:1997ie} study the coupled Einstein-KG equations.  Moffat \cite{Moffat:2005si} and  Moffat  and Toth \cite{Moffat:2014pia} propose TeVeS fields to have a modified gravity.
% see also \cite{Bekenstein:2010pt}.
  Jetzer et al. \cite{Jetzer:1992tog}, Bezares et al.  \cite{Bezares:2022obu}, Liddle and Madsen \cite{Liddle:1992fmk} introduce scalar fields to have boson stars.
    { There are also a large number of papers on regular blackholes  introduced since late 1960s,  \cite{Sakharov:1966aja,Gliner1966cvz} and onwards, see  \cite{Lan:2023cvz} for a short review. We remind that what we do here has no kinship with regular blackholes, where to avoid singularities one resorts to auxiliary energy momentum tensors.
  }

What differentiates this paper from the ones cited above or not is our focus on the fact that particles shaping the structure of the  spacetime are not point-like but extended wave packets.
We find that  due recognition of this wave nature of individual gravitating particles not only removes the essential and coordinate singularities, but alters the spacetime structure at galactic distances. Evidently the cumulative quantum waves  of a coherent and/or random  collection of gravitating particles do not cancel out each other. At far outreaches of a galaxy, say, they create   $r^{-1}$ force and cause  flat rotation curves.   We will come back to this issue in subsection \ref{subsec:rotationcurves} and section \ref{conclusion}.

\section{Definition of the problem}
\label{def.prob}
A pair of neutrons coupled through their isospin behave as an electrically neutral boson of spin $1$ or $0$. See e.g.  \cite{Georgi:2000vve}, chapter 5, section 5.4, Isospin generators . To mimic a neutron star one may consider a collection of such neutron nuggets. Neutrons aside, in a much wider context, one knows that about $99\% $ of the content of the universe is hydrogen and helium. They are either spread out in interstellar and intergalactic spaces or are in the form of ionized particles in an electrically neutral matrix of plasma in the stars. In the absence of any better choice and with some condone we contend to consider the  matter we are dealing with as a collection of $N$ bosons in which each constituent is represented, for the moment, by a  KG   wave.   For the total and individual Lagrangian density of the constituent bosons we write,
\begin{eqnarray}
{{\cal L} }&=& \sum_{i=1}^N {\cal L}_i\,, \label{KGlagtotal}\\
 {\cal L}_i &=& \frac{1}{2}\left(\frac{\hbar^2}{m_i}g^{\mu\nu}\partial_\mu\psi^*(i)\partial_\nu\psi(i)  - m_ic^2\psi^*(i)\psi(i)\right)\,, \label{KGLag}\\
&=& {- {\frac{1}{2}}\psi^*\Big(  \frac{{{\hbar ^2}}}{{{m_i}}}\big({g^{\mu \nu }}{\partial _\mu }{\partial _\nu }\psi (i)\big) + {{m_i}{c^2}\psi (i) }\Big)}\label{KGLag2}\\
 &+&   {\frac{1}{2} \frac{{{\hbar ^2}}}{{{m_i}}}{\nabla _\mu }\Big({g^{\mu \nu }}{\psi ^*}(i){\partial _\nu }\psi (i)\Big) }  \label{tot-derivative}
   \end{eqnarray}
   where $m_i$ is the mass of a single constituent, of the order of one or two nucleon mass.  {The expression in  (\ref{tot-derivative})  is a total derivative and  does not contribute to the equation of motion . The latter from either (\ref{KGLag}) or (\ref{KGLag2}) is}
 \begin{eqnarray}
 \frac{1}{\sqrt{-g}} \partial_\mu \left(\sqrt{-g}\partial^\mu\psi\right) + \frac{m^2 c^2}{\hbar^2}\psi = 0, ~~\textrm{for each boson.}
 	 		  \label{KGeq}
 \end{eqnarray}
 % {Solutions of  \ref{KGeq} for $\psi$ and $\psi^*$ can be complex conjugate of each other or any real combination of them. We will come back to this issue in section \ref{subsecfardistance} where we introduce YKG, the Yukawa ameliorated KG.}
 %
The corresponding energy-momentum tensor of each constituent is,
 \begin{eqnarray}
T_{\mu\nu} &=& - \frac{2}{\sqrt{-g}}\frac{\partial}{\partial g^{\mu\nu}}
\left(\sqrt{-g}{\cal L}\right)  \nonumber\\
&=& -\frac{\hbar^2}{m}\partial_\mu \psi^*\partial_\nu\psi
+ \frac{1}{2} g_{\mu\nu}\left(\frac{\hbar^2}{m}\partial^\alpha\psi^*\partial_\alpha\psi  -mc^2\psi^*\psi\right),  \label{tmunu}\\
T &=& T^\alpha_\alpha = \frac{\hbar^2}{m} \partial^\alpha\psi^*\partial_\alpha\psi
- 2 mc^2\psi^*\psi,                            \label{TKG}\\
\nabla^\nu T_{\mu\nu} &=& 0\,. \label{TKHconserved}
	\end{eqnarray}
 {Note the conservation of the energy-momentum tensor for each gravitating  KG wave.}
  Einstein's field equation for the collection of $N$ boson  may now be written as,
	\begin{eqnarray}		
		R_{\mu\nu} &=&   \frac{8\pi G }{c^4}N \left(T_{\mu\nu} -
	\frac{1}{2} g_{\mu\nu} T\right), \nonumber\\
&=&   \frac{4\pi G Nm}{c^2}\left(g_{\mu\nu} \psi^*\psi
- 2\frac{\hbar^2}{m^2 c^2} \partial_\mu\psi^*\partial_\nu \psi\right)
   \label{ricci}
	\end{eqnarray}
For a standing KG wave of time dependence $\exp(-i\omega t)$, the two terms $\psi^*\psi$ and $\partial_\mu\psi^*\partial_\nu\psi$ are time-independent. This in turn entails time independence of the two coupled and non-linear equations (\ref{KGeq}) and (\ref{ricci}). Thus  one may devise a static and spherically symmetric Schwarzschild-like metric,
\begin{eqnarray}
	ds^2 = -B(r)c^2 dt^2 + A(r) dr^2 + r^2 d\Omega^2, ~~\textrm{signature:} (-,+,+,+). \label{metric}
\end{eqnarray}
Reduction of (\ref{KGeq})  with (\ref{metric})  gives,
\begin{eqnarray}
	\frac{d^2\psi}{dr^2} + \left( \frac{2}{r}
- \frac{1}{2A}\frac{dA}{dr} +  \frac{1}{2B}\frac{dB}{dr}\right) \frac{d\psi}{dr} + \frac{m^2c^2}{\hbar^2}
\left(1 +\frac{1}{B }\frac{\hbar^2\omega^2}{m^2c^4}\r)\psi    = 0. \label{psieq}
\end{eqnarray}
From the following combination of  the components of  (\ref{ricci})  one finds
	\begin{eqnarray}
&& \frac{R_{tt}}{2B} + \frac{R_{rr}}{2A} + \frac{R_{\theta\theta}}{r^2}
	  =  \frac{1}{r^2} \left(\frac{d}{dr}\Big(\frac{r}{A}\Big) -1 \right) \nonumber\\
&& \hspace*{1 cm} = -  \frac{2\pi r_s}C \left(\Big(1- \frac{1}{B} \frac{\hbar^2\omega^2}{m^2c^4}\Big)\psi^*\psi - \frac{1}{A} \frac{\hbar^2}{m^2c^2}
\psi'^* \psi'\right), \label{Aeq}\\
  && 	\frac{R_{\theta\theta}}{r^2} = \frac{1}{r^2}\Big( \frac{1}{A}-1\Big) + \frac{1}{2rA}\frac{d}{dr}\ln\Big(\frac{B}{A}\Big)
	= ~\frac{2\pi N r_s}C  \psi^*\psi.  \label{Beq}\\
 &~&\hspace{1.4 cm} ~ r_s := 2Gm/c^2,  ~~~~ C, \textit{a constant to be discussed}.  \nonumber
 \end{eqnarray}
In (\ref{psieq}) and (\ref{Aeq}),    $(\hbar^2\omega^2/m^2c^4)$  is the square of oscillation energy to the rest mass energy of a  KG wave. And $(\frac{\hbar^2}{m^2c^2} \approx 10^{-31} \textrm{m}^2)$ is the square of the Compton wavelength of the  mass $m$.  Both are extremely small. Unless one is dealing with extreme relativistic cases, one may safely ignore them   without compromising the essential role of the wave nature of the KG field.   Furthermore, it is  preferable to work with dimensionless quantities.  We adopt
$$x = \frac{r}{r_s},~ ~~~~ \frac{d}{dr} = \frac{1}{r_s}\frac{d}{dx}, ~~~~~r_s =2Gm/ c^2. $$
Applying these  simplifications to (\ref{psieq}) and (\ref{Aeq}) we  arrive at.
\begin{eqnarray}
 &&	\psi'' + \left(\frac{2}{x} - \frac{1}{2}\frac{A'}{A} + \frac{1}{2}\frac{B'}{B}\right)\psi'
 	+  k^2 \psi = 0, ~~k = r_s \frac{mc}{\hbar}\ll 1\,,\label{psieqN}\\
 	  \nonumber\\
  	&& \frac{1}{x^2} \left(\frac{d}{dx}\left(\frac{x}{A}\right) -1 \right) =  -	\frac{2\pi N}C  \psi^*\psi\,, \label{AeqN}\\
 	&&  \frac{1}{x^2}\left( \frac{1}{A}-1\right) + \frac{1}{2xA}\frac{d}{dx}\ln\left(\frac{B}{A}\right)  = \frac{2\pi N}C \psi^*\psi\,,  \label{BeqN}
 \end{eqnarray}
where  $(^\prime)$ denotes $d/dx$. The dimensionless $k\ll 1$ is the ratio of the Schwarzschild radius to the Compton wavelength of $m$.

 	\section{Solution by Iteration}\label{sec:solutionbyiteration}
 As an initial guess to begin iteration, we use Schwarzschild's metric coefficients

$$ B_s= \frac{1}{A_s} = \Big(1 - \frac{1}{x}\Big)\,.$$
We substitute them in (\ref{psieqN}) and solve it for $\psi,$ substitute the result in  (\ref{AeqN}) and solve it for $A$, substitute the results in  (\ref{BeqN}) and solve it for $B$.
 In the course of integration in different intervals of $x$, we exercise certain precautions:

 \begin{itemize}
 \item
	Close to the origin, $ (x \rightarrow 0)$, we expand all functions as Taylor series in $x$.
 \item
	Close to the Schwarzschild horizon,  we expand all functions as Taylor series in $(x-1)$.
 \item
	 At far distances, $(x\rightarrow \infty)$, we assume $(B=A^{-1}=1)$.  (\ref{psieqN}) reduces to Helmholtz's equation.  Along with (\ref{AeqN}) and (\ref{BeqN}), they are solved analytically.

\end{itemize}

\subsection{Near origin solutions, $x \rightarrow 0$.}\label{around centre}

Taylor expansion of initial  $A_s$ and $B_s$ in $x$ is
\begin{eqnarray}
  B_s &=& -\frac{1}{x} +1, ~~A_s=-\frac{x}{1-x} =-(x+x^2+\cdots)\,,\nonumber\\
  \frac{B'}{B}\Big|_s &=& -\frac{A'}{A}\Big|_s=- \frac{1}{x (1-x)} = -\frac{1}{x} (1+x+x^2+\cdots)\,. \nonumber
 \end{eqnarray}
Equation  (\ref{psieqN})  reduces   to,
\begin{eqnarray}
&&\psi'' +\left(x^{-1}-1 -x - \cdots \right) \psi'  +  k^2  \psi = 0\,.  \label{psixTendszero}
\end{eqnarray}
From  (\ref{AeqN}), (\ref{BeqN}) and (\ref{psixTendszero}) we now find,
\begin{eqnarray}\label{psi(x)}
&& \psi(x) = 1 - \frac{1}{4}{k^2}{x^2} +\cdots, \\
&& B(x) = \frac{1}{A(x)} = 1 + \frac{2\pi N}{C}\Big( \frac{1}{3}x^2 - \frac{1}{10}k^2  x^4 +\cdots \Big)\,. \label{BAminus1ofx}
\end{eqnarray}
 Essential singularity at the origin is removed. Spacetime is asymptotically flat as $x\rightarrow 0$.

\subsection{Near horizon solution, $x \rightarrow 1$} \label{nearhorizon}

Taylor expansion of   $A_s$ and $B_s$ in $y=x-1, |y| \ll 1$, is
\begin{eqnarray}
B_s &=& \frac{y}{1+y}= y-y^2+y^3+\cdots, ~~~ A_s = y^{-1}+1, \nonumber\\
  \frac{B'}{B}\Big|_s &=& - \frac{A'}{A}\Big|_s = y^{-1} -1 +y -y^2 + \cdots. \nonumber
  \end{eqnarray}
 (\ref{psieqN})  reduces   to,
\begin{eqnarray}
&& \frac{d^2\psi}{d y^2} +\left(y^{-1} + 1 -y  +\cdots \right) \frac{d\psi}{dy}
 +k^2  \psi = 0\,.
\label{psi''horizon}
\end{eqnarray}
To the   order  $y^2 =(x-1)^2$ for $\psi$, and to the order $(x-1)$ for $A$ and $B$ solutions are,
\begin{eqnarray}\label{Psi-horizon}
\psi &=& 1 - \frac{1}{4}k^2 (x-1)^2,   \\
B &=&\frac{1}{A}  = 1+\frac{1}{3}\frac{2\pi N}{C} x^2\left( \Big(1 - \frac{1}{2}k^2\Big) -\frac{3}{4}k^2\Big(x-1\Big)\right).
 \label{BA-horizon}
 \end{eqnarray}
 Unlike the Schwarzschild metric, both $A$ and $B$ keep their spacelike and timelike nature and are continuous before and after $(x =1)$. There is hardly any justification to use the nomenclature  \textit{'horizon' }.
 It is still, however,  conceivable to have non-singular blackholes. This may happen if at least a fraction of the gravitating KG fields goes into  bosonic degenerate states. Heisenberg's uncertainty, however, will not allow   gravitational forces to crush the degenerate matter to singularity.\\

 { At this stage it is appropriate to pay tribute to Roy Kerr.
In a recent paper \cite{Kerr:2023rpn} he expresses reservations    on the singularity theorems of  Penrose \cite{Penrose:1964wq} and of Hawking \cite{Hawking:1971vc}:  \textit{"there are no proofs that black holes contain singularities when they are generated by  real physical bodies".}
In his concluding remarks Roy further  reiterates:
\textit{"The  author has no doubt that, and  he  never  did, that when Relativity and Quantum Mechanics are melded it will be shown  that there are no singularities anywhere."}\\
\\
What we conclude above is in agreement with  Roy Kerr's prophecy.
Evidently the sole replacement of point-like gravitating masses by wave like ones is sufficient to wipe out  both intrinsic and coordinate singularities from GR.
}

 \subsection{Far distance solutions, $\mathrm{x \rightarrow \infty}$}\label{subsecfardistance}

 As $x$ tends to infinity   $A ~\textrm{and} ~B \rightarrow 1$.  (\ref{psieqN}) reduces to the standard Helmholtz equation,
 \begin{eqnarray}
 \psi'' + \frac{2}{x}\psi' + k^2 \psi =0\,.\label{psi.infty}
 \end{eqnarray}
 The lowest order solutions of (\ref{psi.infty})  are   $\frac{1}{x} e^{\pm ikx}$.  None of these solutions,  however, are normalizable, for as $x\rightarrow \infty$. The wave amplitude does not falloff steeply enough to have finite $\int\psi^*\psi$. In the near origin and near horizon solutions we were allowed to ignore the inconsistency,  for divergence of the normalization constant concerned far distance behaviour of the KG wave.

 To remedy the case we suggest a Yukawa-ameliorated Klein-Gordon (YKG) wave, $\psi =\frac{1}{x} e^{-(\kappa\pm i k)x}$, where we will later find that $\kappa$ is an exceedingly small positive number. The wave equation (\ref{psi.infty}) changes accordingly,
 \begin{eqnarray}
 && \psi'' +\frac{2}{x} \psi' - (\kappa\pm ik)^2 \psi = 0\,. \label{psiYKG}
 \end{eqnarray}
 { This equation can be derived from:}
\begin{eqnarray}
 {{\cal L}_i(YKG)}&=& { - {\frac{1}{2}}\psi^*\Big(  \big({g^{\mu \nu }}{\partial _\mu }{\partial _\nu }\psi (i)\big)} -  {(\kappa\pm i k)^2 (i)\psi (i) \Big)}\,.\label{KGLag2-YKG}
   \end{eqnarray}
 { The Lagrangian  (\ref{KGLag2-YKG}) is the same as that in (\ref{KGLag2}), simplified to account for the flatness of space-time and generalized to accommodate the Yukawa falloff parameter $\kappa$.
}
 Such Yukawa amelioration  is a logical and perhaps the imperative way out of the dilemma, for it is inconceivable to imagine that the wave function of a single particle extends to infinity and alters the spacetime structure at the scale of Universe.

The two lowest order solutions of \eqref{psiYKG} are $x^{-1}e^{-(\kappa\pm ik)x}$. Any linear combination of them is a legitimate solution. We choose the following real and normalized combination,
\begin{eqnarray}
&& \psi = \frac{1}{\sqrt{C}}\frac{1}{x}e^{-\kappa x}\cos(kx+\alpha)\,,  ~~ ~~  0\leq \alpha \leq \pi/2\,,~~\textit{ a mixing parameter.}\label{psiYKGreal}\\
&& C =\int_0^\infty \frac{1}{x^2}e^{-2\kappa x}\cos^2(kx+\alpha) d^3 x = \pi \Big(\frac{1}{\kappa}
+ \frac{\kappa \cos 2\alpha -  k \sin 2\alpha}{\kappa^2  + k^2} \Big)\simeq\frac{\pi}{\kappa}\,, \nonumber
\end{eqnarray}
   Substitution of (\ref{psiYKGreal})   in (\ref{AeqN}) and (\ref{BeqN}) gives.
     \begin{eqnarray}
 &&  A^{-1} =1 - \frac{N}{{2x}}{e^{ - 2\kappa x}}\Big(1 - \frac{\kappa }{k}\sin 2(kx + \alpha )\Big)\,,    \label{Aminus1infty}\\
 &&    A =1 + \frac{N}{{2x}}{e^{ - 2\kappa x}}\Big(1 - \frac{\kappa }{k}\sin 2(kx + \alpha )\Big)\,,    \label{Ainfty}\\
 &&    B =1 - \frac{N}{{2x}}{e^{ - 2\kappa x}}\Big(1 + \frac{3}{4}\frac{\kappa }{k}\sin 2(kx + \alpha )\Big)\,.  \label{Binfty}
    \end{eqnarray}
     In the weak field approximation the gravitational force is
       \begin{eqnarray}
      	\frac{1}{2} c^2  B' &=& N c^2 e^{ - 2\kappa x}
       \bigg( \frac{\kappa}{x}\Big(1  - \frac{3}{4} \cos 2(kx + \alpha)  \Big)  \nonumber\\
       &~& \hspace*{2cm}    + \frac{1}{2 x^2}\Big(1 + \frac{3}{4}\frac{\kappa}{k}\sin 2 (kx +\alpha) \Big) \bigg),                                          \label{grav.force}
       \end{eqnarray}
 where terms of order $\kappa^2$ are omitted. As $x\rightarrow \infty$:  a)  The spacetime becomes asymptotically flat. b) The  gravitational force becomes  dominantly $x^{-1}$ rather than the Newtonian $x^{-2}$. c) There are  sinusoidal variations  in \eqref{Ainfty}-\eqref{grav.force}. We will come back to the role of  these features shortly.

 \subsection{Time to revise our primitive model}

  In Section \ref{def.prob} we assumed the gravitating body consists of a collection of $N$ bosons, each  represented by a KG, or now by a YKG, field. All constituents  were assumed to be in their  ground states. This is perhaps a good approximation for the  core of those galaxies that have a central blackhole whose degenerate constituents are arranged in an orderly manner in the lowest quantum states in a coherent and phase-tuned manner (i.e. the same $\alpha$).
  The much larger fraction of the gravitating body, however, remains  non-degenerate. Its constituents reside in the spacious phase space of states with no obligation to tune themselves with neighbors. Therefore, let us rewrite the Lagrangian of (\ref{KGLag}) as the sum of two degenerate and non-degenerate components:         %
\begin{eqnarray}
{\cal L} =  (fN)  {\cal L}_{deg}(ground. state) + \sum_i  ^{(1-f)N} {\cal L}_{non.deg}(i)\,, \label{KG.Lag.revised}
\end{eqnarray}
where we have divided the gravitating matter into a fraction $f$ of degenerate core in their ground state and the remaining non-degenerate fraction $(1-f)$ in various $\psi(i)$ states
 (see \cite{Davis_2019}  for blackholes at the center of spiral galaxies   and the references therein. Typical values of $f$ are $ 10^{-3} - 10^{-4}$).
   Constituents in the non-degenerate fraction will necessarily be in different non-correlated states and for all practical purposes will have  random  phases, $\alpha_i$. In line with this division of $\cal L$ the metric coefficients $A$ and  $B$ divide  accordingly.   {For instance $A$ of \eqref{Ainfty} gets divided as follows
   \begin{eqnarray}\label{Ainfty0}\nonumber
A(x)& =& 1 + f\frac{N}{{2x}}{e^{ - 2\kappa x}}\Big(1 - \frac{\kappa }{k}\sin 2(kx + \alpha )\Big)\\
&+& \sum\limits_i^{(1 - f)N} {\frac{1}{{\left| {\mathbf{x} - {\mathbf{x}_i}} \right|}}} {e^{ - 2\kappa \left| {\mathbf{x} - {\mathbf{x}_i}} \right|}}\Big(1 - \frac{\kappa }{k}\sin 2(k\left| {\mathbf{x} - {\mathbf{x}_i}} \right| + {\alpha _i})\Big),
   \end{eqnarray}}
    where $\textbf{x}$ and $\textbf{x}_i$ are coordinates of the observation point and the source ones, respectively.
Both $\textbf{x}_i$ and $\alpha_i$ are random variables and $(1-f)N$ is a huge number.  At far distances,
($|\mathbf{x}| \gg |\mathbf{x}_i|, \forall |\mathbf{x}_i| $), one may Taylor expand  terms of the form $f(|\mathbf{x}-\mathbf{x}_i|)$ as follows
\begin{eqnarray}
 f(|\mathbf{x}-\mathbf{x}_i|) = f(x) - \frac{df}{dx }\frac{\mathbf{x}.\mathbf{x}_i}{x^2}
+\frac{1}{2}\frac{d^2 f}{dx^2}\Big(\frac{\mathbf{x}.\mathbf{x}_i}{x^2} \Big)^2 + \cdots \,. \nonumber
\end{eqnarray}
Summing over $i$ eliminates all terms odd in $x_i$ due to their random nature. The next even term is an order of magnitude $(|\mathbf x_i|/|\mathbf x|)^{2}$ smaller than the first term and  can be dropped as $x\rightarrow 0$ (monopole approximation).
With the two trigonometric terms in  (\ref{Ainfty0}), however, one should be careful.  They  are high-frequency random undulations. The routine to deal with them is to square random terms, sum them up, and take the square root of the sum.  {Reducing (\ref{Ainfty0})  as explained gives,
   \begin{eqnarray}\label{Ainfty01}\nonumber
A(x) &= &1 + f\frac{N}{{2x}}{e^{ - 2\kappa x}}\Big(1 - \frac{\kappa }{k}\sin 2(kx + \alpha )\Big)\\
&+& (1 - f)\frac{N}{{2x}}{e^{ - 2\kappa x}} - \frac{\kappa }{{4kx}}\sqrt {(1 - f)N} {e^{ - 2\kappa x}}\,,
   \end{eqnarray}
   as noted above the factor $(1-f)N$ is a huge number. One may drop its  square root and arrive at
   \begin{equation}\label{Ainfty02}
A(x) = 1 + \frac{N}{{2x}}{e^{ - 2\kappa x}}\l(1 - f\frac{\kappa }{k}\sin 2(kx + \alpha )\r)\,.
   \end{equation}
For the  $B$ of \eqref{Binfty}, one similarly finds
\begin{eqnarray}\label{Binfity0}\nonumber
B(x) &=& 1 + 2\frac{{\Phi (x)}}{{{c^2}}}\\
&=& 1 - \frac{N}{{2x}}{e^{ - 2\kappa x}}\l(1 + \frac{{3\kappa }}{{4k}}f\sin 2(kx + \alpha )\r)\,,
\end{eqnarray}
where in the weak field approximation,
\begin{eqnarray}\label{gpotential}\nonumber
\Phi (x)&=&\Phi (r/r_s)=\frac{1}{2}c^2(B-1)\\
&=&- \frac{Nc^2}{{4x}}{e^{ - 2\kappa x}}\l(1 + \frac{{3\kappa }}{{4k}}f\sin 2(kx + \alpha )\r),
 \end{eqnarray}
  is effectively the gravitational potential.
Accordingly the gravitational force field becomes
\begin{eqnarray}\label{gravforceinfty}\nonumber
g(r) &=&  - \frac{1}{2}{c^2}\frac{{dB}}{{dr}} =  - \frac{1}{2}\frac{{{c^2}}}{{{r_s}}}\frac{{dB}}{{dx}}\\\nonumber
&\approx& \frac{1}{2}\frac{{{c^2}}}{{{r_s}}}{e^{ - 2\kappa r/r_{s}}}\l(\frac{N}{{2{x^2}}} + \frac{{2N\kappa }}{x}\big(1+\frac{3}{2}f\cos2(kx+\alpha)\big)\r),\\
& \approx& {e^{ - 2\kappa r/r_{s}}}\l(\frac{{GNm}}{{{r^2}}} + \frac{{N{c^2}\kappa }}{r}\big(1+\frac{3}{2}f\cos2(kr/r_s+\alpha)\big)\r).
\end{eqnarray}
}
 {Ignoring the exponential  factor in \eqref{gravforceinfty}, the $\frac{1}{r^2}$ term is the familiar Newtonian gravitational force. The $\frac{1}{r}$ term is non-Newtonian. It is the collective quantum effect of $N$ bosons in the model ($N=10^{69}$ nucleons in a  Milky Way type galaxy). It falls off at much slower rate than the Newtonian force and is essentially responsible for  flat rotation curves in actual spiral galaxies. The sinusoidal factor in \eqref{gravforceinfty}, proportional to $f$, comes from a would be central blackhole.  }

\subsection{Determination of $\kappa$ - Tully- Fisher relation}\label{subsecTFR}
\begin{figure}[h]
	\centering
\includegraphics[width=0.42\linewidth]{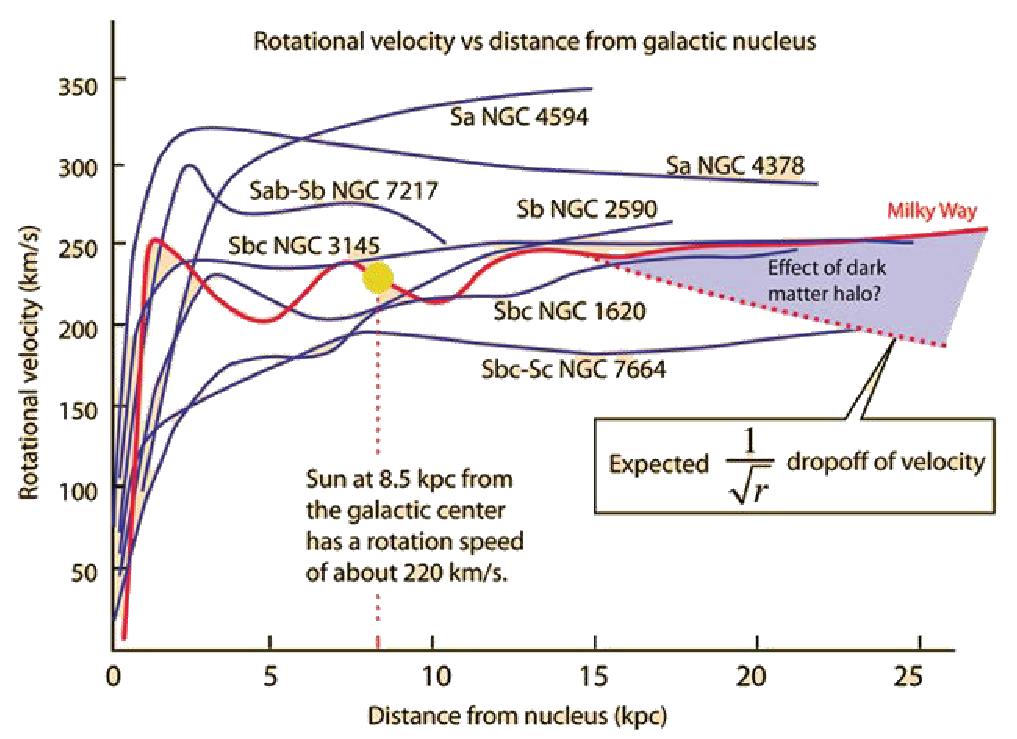}
	\caption{A sample of rotation curves of spiral galaxies - Carroll and Ostlie, Ch 24 \cite{Carroll-17}; Wilson et al., \cite{Rholf};  Freedman and  Kaufmann \cite{Kaufmann}}
	\label{fig:rotationcurve00}
\end{figure}

\textbf{	Observed facts and implication: }
In Newtonian gravitation  the speed of a test object in circular orbit about a localized gravitating mass is
$v^2=\frac{1}{2} c^2 r\frac{dB}{dr} = \frac{GM}{r}   $.  This, however,  is not the case  for stars and HI clouds orbiting their host   galaxies at large distances. The Tully-Fisher relation (TFr), sifted from a forage  of observational  data, \cite{Tully:1977fu}, states that:
 \begin{itemize}
 \item
  	The observed rotation speeds, $v^2$,  of distant stars and neutral hydrogen clouds  in spiral galaxies show a much gentler decline, if any, compared to the expected $GM/r$ falloff predicted by the Newtonian gravitation,
  see Figs. (\ref{fig:rotationcurve00}) and (\ref{fig:screenshot001}).
 \item
  Beyond the visible disks of galaxies, the observed  $v^2$  has , more often than not,  the flat asymptote $v^2 \propto \sqrt{M}$, rather than $\propto M$.
See  \cite{Sobouti:2008xz,shostak1973aperture,Bosma:1981zz,Begeman:1989kf,Begeman:1991iy,Sanders:1998gr,Sanders:2002pf,Persic:1995tc,Persic:1995ru}.
 \end{itemize}

In his Modified Newtonian Gravity (MOND), Milgrom interprets TFr by saying that the effective gravitational acceleration in galactic scales is,
\begin{eqnarray}
g_{eff}= g_{New} ~~\textit{if large}, ~~~ = \sqrt{g_{New}a_0}, ~~\textit{ if small},~~ a_0= 1.2\times 10^{-10} \textit{m.sec}^{-2}\,. \nonumber
\end{eqnarray}
  Milgrom is silent as to whether his $a_0$ is a universal acceleration applicable to all spirals or not. There are, however, indications  that $a_0$ might depend on the size and/or mass of the host galaxies, see \cite{Sobouti:2006rd} for $a_0$ in a list of 53 spiral galaxies.

 % \subsection{ Determination of $\kappa$ }\label{subseckappa}

Leveraging the Tully-Fisher Relation (TFr) and MOND, we can now determine the value of $\kappa$. Of the three $\kappa/r$ terms in (\ref{Binfity0}), the one with  factor $\sqrt{N}$ is small and can be neglected. The sum of the remaining two (again ignoring the falloff factor and $(\kappa,~ k)$ corrections) can be  written as follows,
\begin{eqnarray}
Nc^2 \frac{\kappa}{r} = \sqrt{\frac{N Gm}{r^2}\Big(\frac{Nc^4\kappa^2}{Gm}\Big)}
= \sqrt{g_{New}\Big(\frac{Nc^4\kappa^2}{Gm}}\Big)\,.\nonumber
\end{eqnarray}
It only suffices to identify $\Big(\frac{Nc^4\kappa^2}{Gm}\Big)$ with Milgrom's $a_0$ and obtain,
\begin{eqnarray}
\kappa =  \sqrt{\frac{Gm}{Nc^4}a_0} \approx 4.06\times 10^{-41}N^{-1/2}\,. \label{kappa}
 \end{eqnarray}
 %
% Note the dependence  of $\kappa$  on $N^{-1/2}$, which means dependence on the inverse square root of the galactic mass.
We recall that  $N$   is almost  the number of nucleons in the galactic matter. For the  Milky Way   of about $1.5\times 10^{12}M_\odot$, one has
 \begin{eqnarray}
 && N_{MW} \approx 1.7\times 10^{69}, ~~~~\sqrt{N_{MW}} \approx 4.2\times 10^{34}\,. ~~~~
\kappa_{MW}\approx 9.6\times 10^{-76}, \nonumber
 \end{eqnarray}
 Note how small $\kappa$ an how gentle the exponential falloff  factor, $e^{-2\kappa x}$,  are.\\

\subsection{Rotation curves}\label{subsec:rotationcurves}

\begin{figure}[h]
	\centering
\includegraphics[width=0.45\linewidth]{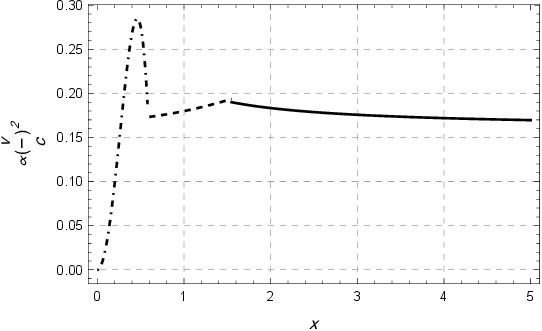}
\includegraphics[width=0.46\linewidth]{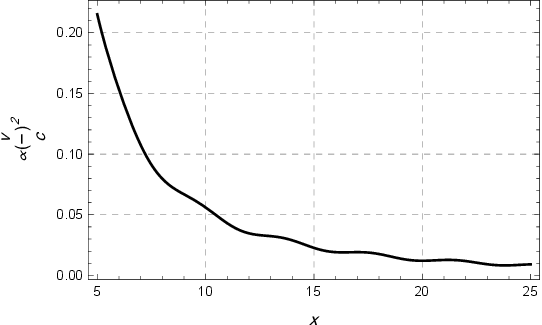}
	\caption{ {Rotation curve of a  spherically symmetric toy galaxy. It represents an initial steep  rise to a maximum followed by a gentle decline towards an almost flat asymptote.
The model does not account for the complex structures of real galaxies, such as galactic planes or spiral arms. Even the central mass is approximated by its monopole moment.
In the right hand panel we have scaled up the periodic undulations. They are caused by  the phase-tuned degenerate core of the model and  are quantum effects.}\\
 \\
Free parameters: $N=10^{70}$, $\kappa=10^{-3}$, $k=0.2$, $f=10^{-2}$, $\alpha=0$.}
	\label{fig:RotationCurve0ursToyGalaxy}
\end{figure}

\begin{figure}[h]
	\centering
	
\includegraphics[width=0.43\linewidth,height=0.36\linewidth]{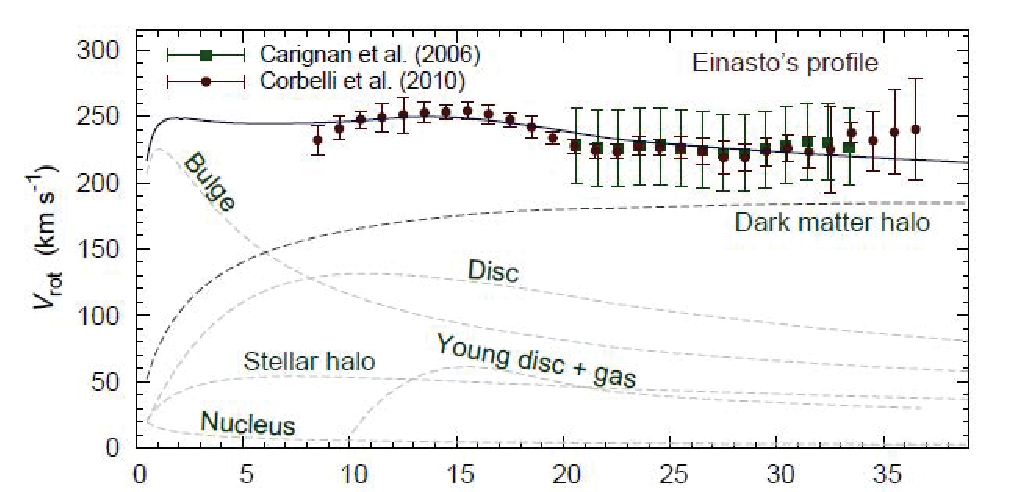}
	\includegraphics[width=0.45\linewidth,height=0.37\linewidth]{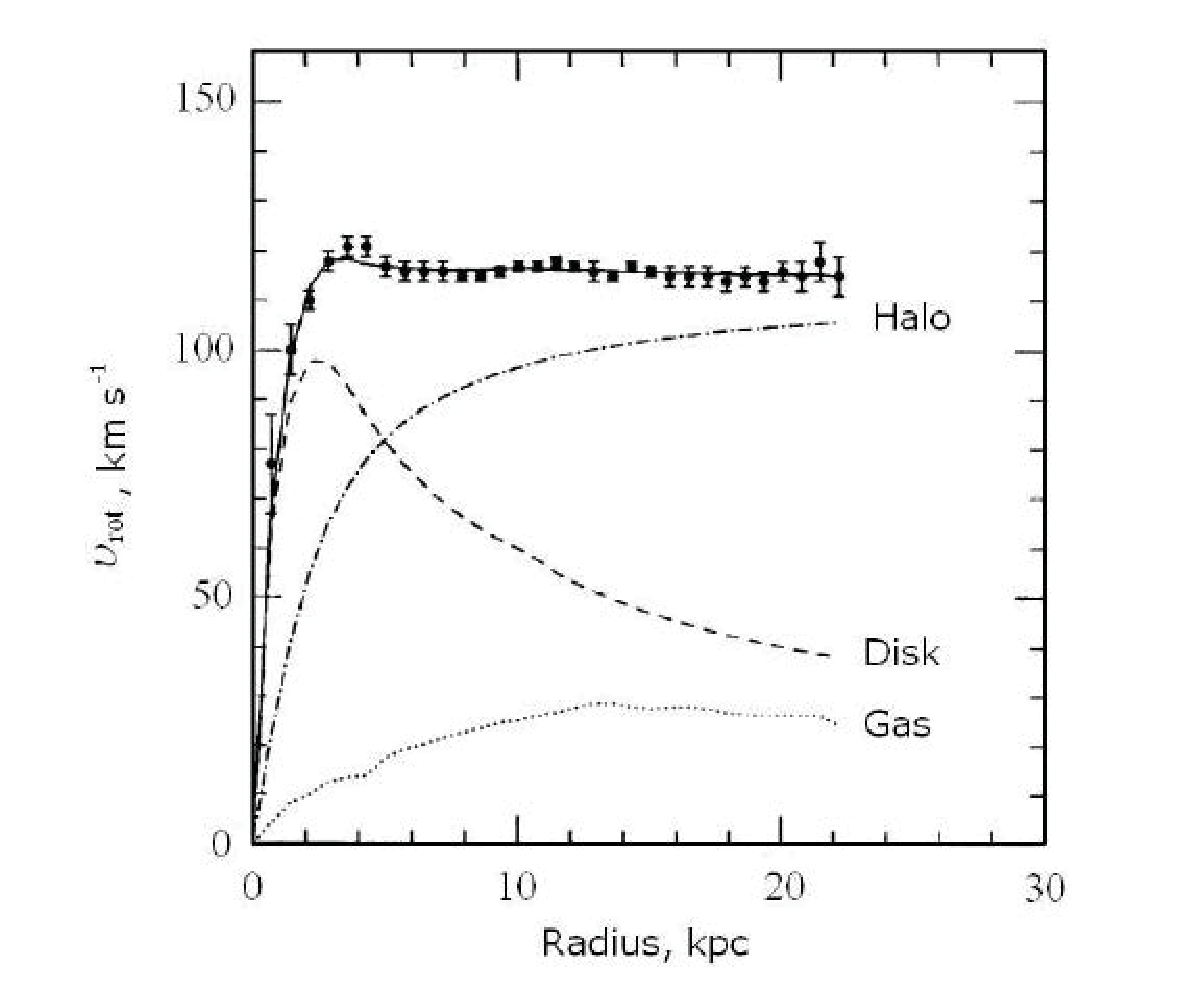}
	\caption{Left panel: Rotation curves of M31 - Reference: Tamm et al.,	Astronomy \& Astrophysics 546,{2}, A4, 2012.
Right panel: Rotation curve of NGC 6503-Reference: Freese K, arXiv:0812.4005, \cite{Freese:2008cz}.\\
 Note  periodic undulations on the observed dotted points and compare them with those in Fig (\ref{fig:RotationCurve0ursToyGalaxy}). }
	\label{fig:screenshot001}
\end{figure}

Having calculated  $A$ and $B$ for a hypothetical galaxy consisting of a collection of KG and/or YKG gravitating bosons, we are now in a position to analyze  trajectories of test objects in circular orbits around the  center. Instead of analyzing    geodesic equations, which is the systematic way of carrying out the task, we take a short cut. In a weak gravitation the circular speed is given by,
 {$$ v^2 =  r g(r)={e^{ - 2\kappa r/r_{s}}}\l(\frac{{GNm}}{{{r}}} + {{N{c^2}\kappa }}\Big(1+\frac{3}{2}f\cos2(kr/r_s+\alpha)\Big)\r).$$
 Fig.(\ref{fig:RotationCurve0ursToyGalaxy}) is the rotation curve of our toy galaxy.
   We consider a spherically symmetric collection of YKG gravitating waves  in its monopole approximation. There are no  galactic planes  or spiral arms in the model to mimic  any realistic spiral or non-spiral  galaxy. All we want to demonstrate is that  moving outward from the center, there is steep rise to a maximum followed by a gentle declines towards an almost flat asymptote.}

The free parameters of the model are $N = 10^{70}, \,\kappa = 10^{-3}, \,k = 0.2,  \,f = 0.001\,,\alpha=0.$
 The model has a central blackholes of  $0.001$th   of the galaxy. Superimposed on the almost flat branch of the curve, there are  short-wavelength and small-amplitude sinusoidal variations.
 They come from the degenerate and phase-tuned  matter of the central blackhole.  Compare them with similar variations  in  Fig.(\ref{fig:screenshot001}) for the rotation curves of M31 and NGC 6503.
 To the best of our knowledge they are never before, addressed in the astronomical literature.
 See also Fig.(\ref{fig:rotationcurve00}), for assorted rotation curves from  Carroll and Ostlie \cite{Carroll-17}; Wilson et al. \cite{Rholf};  Freedman and  Kaufmann \cite{Kaufmann}.

\section{Massive Gravitons}\label{sucsec:massivgraviton}

\textbf{ A Historical Note}: Yukawa  introduced his field in 1935 to explain the residual strong interactions that bind  nucleons together within a nucleus at femtometer scales. The mediating bosons, $\pi^\pm$, were discovered in 1947 and $\pi^0$ in 1950. There are two masses associated with Yukawa paradigm:   a) the masses  of  interacting quarks between nucleons and b)  the masses of the mediating pions.

Our YKG is the gravitational analog of the Yukawa field, albeit operating on galactic scales.
There are also two masses in YKG: a) the mass of the gravitating nucleons  incorporated in
$k = r_s\frac{mc}{\hbar}=\frac{2Gm^2}{c\hbar}$, (\ref{psieqN}); and b) the   mass of the mediating graviton.  The Compton mass and wavelength associated with $\kappa$ are,
\begin{eqnarray}
&& \lambda_g = \frac{r_s}{2\kappa}  =\frac{Gm}{c^2\kappa}=3.04\times 10^{-14}
 \left(\frac{M}{m}\right)^{1/2} \textrm{meter},\label{graviton.wl}\\
 &&    m_g(\lambda_g) = \frac{\hbar}{c\lambda_g} =1.16\times 10^{-29}\left(\frac{M}{m}\right)^{-1/2}~ \textrm{kg}=6.49\times 10^4 \left(\frac{M}{m}\right)^{-1/2} \textrm{eV}\,.
 \label{graviton.mass}
\end{eqnarray}
 Note the proportionality of  Compton parameters of our  graviton on the square root of the mass of the host gravitating body. The heavier the host the longer the Compton wavelength, and the lighter  the graviton.

  For a Milky Way type galaxy  one gets
\begin{eqnarray}
&&\lambda_g({MW}) \approx 1.2 \times 10^{21} \textrm{m} \approx 38~ \textrm{kpc},\label{MWgravWl}\\
\hspace{.5 cm}
&& m_g({MW}) \approx 2.8 \times  10^{-64}~ \textrm{kg} \approx   9.5\times 10^{-29}~  \textrm{ eV}.
\label{MWgravMass}
\end{eqnarray}

\noindent
\textbf{Two  Features to note:}

 a) Transition  from the Newtonian to the non-Newtonian regime takes place, where the two forces become equal,
\begin{eqnarray}
\frac{GNm}{r_{tr}^2}=\frac{Nc^2\kappa}{r_{tr}}=\sqrt{\frac{GNm}{r_{tr}^2}a_0}
\Longrightarrow~ r_{tr} = \frac{Gm}{ c^2\kappa} = \lambda_g=r_s/2\kappa. \label{r_transition}
\end{eqnarray}

b)  At the same  $r_{tr}= \lambda_g$   the exponential  falloff   reduces to $e^{-1}$, an indication of the gentleness of the Yukawa falloff factor.\\

 {
 The $m_g(r_{tr})$ of \eqref{graviton.mass} mediates the  YKG gravitation at the location $r_{tr}=\lambda_g.$ One expects as one moves from $r_{tr}$ to a different location $r$ this mass to change to $m_g(r),$ but how?   To answer the question we note that   the $m_g(r_{tr})$ does not have kinetic energy, for the  YKG gravitation is static.  It does not have  a rest mass energy either. There only  remains to be a potential energy at $r_{tr}$. To see this, let us multiply $m_g(r_{tr})$ by $e^{-2\kappa r_{tr}/r_s} $  and divide by the same
  $$m_g(r_{tr})=\frac{\hbar}{c r_{tr}}e^{-2\kappa r_{tr}/r_s}e^{2\kappa r_{tr}/r_s}.$$
  We recognize that ${\hbar}/{c r_{tr}}e^{-2\kappa r_{tr}/r_s}$ is proportional to the gravitational potential at $r_{tr}$ and  conclude that upon  moving  from $r_{tr}$ to a generic $r$  to have
  \begin{equation}\label{mgofr0}
m_g(r)=\frac{\hbar}{cr}e^{-2\kappa (r-r_{tr})/r_s}.
\end{equation}
We will come back to this \eqref{mgofr0}, when we discuss the graviton deduced from the orbital precession of $S2$ star orbiting the  supermassive blackhole in  the Constellation Sagittarius A* (SMBH SgrA*) .
}\\

 In Table \eqref{table:1} we have calculated the  mass and the wavelength of  gravitons for the Milky Way, the supper massive blackhole Sgr A*, the Sun, the Earth,  one kg mass, nucleons, and $d$ and $u$ quarks
\begin{table}[h]
 \tbl{Compton wavelength and mass of massive gravitons that mediate the YKG gravitation of a wide range of masses, from celestial bodies to quarks. }
{\begin{tabular}{|cc|c|c|cc|}
\hline
%\centering
%
&      &Mass (kg)& $\lambda_g(m)$&$m_g$ & \\
\hline
&MW&$ 1.5 \times 10^{42}$&$ 1.2\times10^{21}\approx38~\textrm{kpc}$&$ 9.5\times 10^{-29}$eV &\\
  \hline
 & Sun& $2\times 10^{30}$& $1.06 \times 10^{15}\approx7100~\textrm{AU}$&$1.7 \times 10^{-22}$ eV&\\
  \hline
  &SMBH SgrA*& $8.6\times 10 ^{36}$& $1.47\times 10^7$AU= $71$ ps&
$8\times 10^{-26}$ eV&\\
\hline
  & Earth& $ 6\times 10^{24}$ &  $1.7 \times 10^{12}\approx11.3~\textrm{AU}$& $1.06 \times 10^{-19}$eV&\\
 \hline
 &  1 kg& $ 1$ & $0.74$ & $26 \times 10^{-6}$eV &\\
 \hline \hline
 &p \& n & $ 1.673\times 10^{-27}$ &
  $ 30\times10^{-15}$&
  $ 6.4\textrm{ MeV}$ & \\
  \hline
  &  d quark & $8.54\times 10^{-30}$ &  $2.2\times 10^{-15}$ & $85.8 \textrm{ MeV}$ &\\
 \hline
 &u quark& $ 3.56 \times 10^{-30}$ & $1.4\times 10^{-15}$& $134.6 \textrm{ MeV}$ & \\
 \hline
 \end{tabular}\label{table:1} }

\end{table}

  \begin{itemize}
 \item
      In the Milky Way, transition from   Newtonian  to the non-Newtonian gravitation  takes place at $38$ kpc, far outside of the visible disc of the galaxy of radius $26.8$ kpc.
  \item
   {The case of SMBH Sgr A* will be discussed below.}
  \item
       For the Sun (or rather  the whole solar system)  transition takes place at $7100$ AU, somewhere within  the Oort clouds,  believed to have an inner edge  of $2000$ to $5000$ AU and an outer edge of $10000$  to $100000$ AU, see https://science.nasa.gov/solar-system/oort-cloud.
     \item
            For the Earth transition distance is  longer than  Jupiter's orbital radius,  9.54 AU.
      \item
            For a one kg weight,  $\lambda_g = 74$ cm is small enough to think of a table top setup to  check deviations from  Newtonian gravitation.
      \item
             The last four entries for  $ p, n,  d,  u $  are included as a matter of curiosity. They should not be taken  seriously at this stage.
              One, however, wonders  how and why the gravitons  mediating  the gravitational interaction of  $u$ and $d$,  and pions mediating the residual strong interaction of  the same $u$ and $d$ have almost identical masses, $\simeq 130 -140 ~\textrm{MeV}$. There might be a deeper connection between  the YKG gravity and, at least Yukawa's residual, strong interactions.

 \end{itemize}

 {The SMBH SgrA* at the center of the  Milky Way has a mass of $4.3\times 10^6 M_\odot=8.6\times 10^{36} kg$. The S2 star, a bright  infrared source of radiation,   orbits SMBH SgrA* with the eccentricity $e=0.885$ and semi major axis $970 AU$. There are almost 30 years of observational data on the S2 orbit. In
\cite{Zakharov:2016lzv,Jovanovic:2023tcc} the authors use  a Yukawa modified gravity of their own, calculate  the periastron precession of the orbit  and report a graviton mass of $m_g(S2)\leq 2.9 \times 10^{-21}eV$.}

 {To compare this graviton mass with the one reported in Table  \ref{table:1}  ($8\times10^{-26}$  eV)
let us note the former is  obtained from the analysis of  the S2 orbit. The  rough distance of S2 from SMBH SgrA* is about $\approx700 AU$, the sum of the semi major + semi minor axes divided by 2.
The graviton mass we report belongs to a distance $r_{tr}=\lambda_g = 1.47\times 10^{7}AU$. Considering  \eqref{mgofr0},  we have
\begin{eqnarray}\nonumber
\begin{aligned}
{m_g}(700\,AU) =
{m_g}(1.47 \times {10^7}\,AU)\frac{{1.47 \times {{10}^7}}}{{700}}{e^{ - 2\kappa (1.47 \times {{10}^7} - 970)/{r_s}}}
 \simeq 1.68 \times {10^{ - 21}}eV.
 \end{aligned}
\end{eqnarray}
 This number is indeed  in excellent agreement with $m_g(S2)\leq 2.9 \times 10^{-21}eV$ reported in \cite{Zakharov:2016lzv,Jovanovic:2023tcc}.
}

In addition to the massive graviton of SMBH SgrA*, there are further literature on  the mass of graviton depending on how the assumed gravitation differs from the standard GR \cite{Isham:1971gm,Deser:1981wh,deRham:2010kj,deRham:2014zqa,Arkani-Hamed:2002bjr}, or how they are deduced from astronomical observations.
 {The case of  SMBH SgrA* was just discussed and we found in good agreement with what we find. Another closest to what we propose here is \cite{Desai:2018swo}. There a Yukawa-ameliorated gravitation is used to deduce the mass of graviton from observation on Abell1689 galaxy cluster. They report
\begin{equation}\label{MAbell}
 m_g({Abell 1689}) < 1.37\times 10^{-29} ~ \textrm{eV}\,.\nonumber
\end{equation}}
This graviton  mass is a factor of about $7$ lighter than our $ 9.5 \times 10^{-29}$  eV  for the Milky Way.  This means that the graviton from Abell 1689 reaching   earth-bound  observers, has trespassed  a mass of $7^2=49$  Milky Ways.
  Abell 1689  galaxy cluster is reported to have about 1000 galaxies,  \cite{Abell:1989mu}.
 A recent paper \cite{DAgostino:2024ojs}, employs a Yukawa cosmology to constrain the graviton mass from observations of the Milky Way and M31. Although their approach and  reasoning differs from ours, the fact both they and we speak of Yukawa in galactic and cosmic contexts, conclusions are very  much the same. For instance, both studies address  deduction of graviton mass from observations of rotation curves, or both maintain rotation curves should be considered as quantum manifestations of galactic matters. Their graviton mass $m_g\approx1.54\times 10^{-62}$ kg  $= 8.64 \times 10^{-27}$ eV is,  however, is $91$ times larger than our, $9.5 \times 10^{-29}$ eV for the Milky Way.

\section{Conclusion}\label{conclusion}

We argue that matter in classical  physics, including GR, is a collection of discrete point-like entities, to which in defiance of Heisenberg's principle of uncertainty, one assigns precise coordinates and momenta. Singularities and divergences  encountered in Schwarzschild's solution  or elsewhere in GR stem from this non-quantum act.

We propose to replace the discrete collection of point-like masses of GR by a  collection of  wave-like entities,  KG and/or YKG waves.
Waves, on the one hand, are spatially distributed entities and  do not end up in spacetime singularities, as one approaches the origin.  Waves,
on the other hand, reach infinity and produce  $\frac{1}{r}$  forces. The latter  feature alone is capable of explaining  the flat rotation curves of spiral galaxies,  two birds with one stone.

The far zone solutions are the highlight of this paper.  YKG waves reach infinity as $\frac{1}{r}\exp(-\kappa/r) \cos(kr +\alpha)$ if  degenerate, and as $\frac{1}{r}\exp(-\kappa/r)$ if non-degenerate, see \eqref{psiYKGreal}, \eqref{Ainfty} and \eqref{Binfty} . At intermediate distances, the gravitational force is dominantly $\frac{1}{r^2}$, but transits to  $\frac{1}{r}$ as  $ r \rightarrow \infty$.  The latter is responsible for the flat rotation curves of spirals and  can be interpreted as an  alternative gravity, a dark matter scenario, or whatever.

There are ample astronomical observations to support our propositions and conclusions.
In Fig.(\ref{fig:RotationCurve0ursToyGalaxy})   of our toy rotation curve there are small sinusoidal variations on the asymptotic branch. They are there because we have chosen  real YKG waves,
\eqref{psiYKG}, and have assumed that a fraction $f$ of them  are in  degenerate and phase-tuned states. There are frequent observational examples of  such periodic variations, see e.g.  Fig. (\ref{fig:screenshot001}) for the rotation curves of M 31 and NGC 6503. See also Fig. (\ref{fig:rotationcurve00}) for an assorted rotation curves given by Caroll et.al.

Nowadays quantum  technologists are sending and receiving  quantum messages across distances of tens and hundreds of kilometers through the science and technology of quantum entanglement. Do astronomers have sufficient evidence to claim they are observing quantum effects across galaxies through rotation curves? We think they do;  we maintain that  the   Tully-Fisher relation is the cumulative  quantum relic of some $10^{70}$ nucleons in  galactic matters.

In the standard GR gravitation is mediated by massless gravitons and are known to come in two polarizations. The YKG gravitons are massive and (we will show in a forthcoming communication that) they have at least three polarizations. They can be put  in one-to-one correspondence with the three Yukawa pions that mediate the residual strong interactions. There are further similarities between YKG and the residual strong interactions. For instance, both are attractive forces. Or as shown in Table (\ref{table:1}), the mass of YKG graviton, for the gravitational interaction of nucleons, and the $d$ and $u$ quarks are nearly identical to the masses of $\pi$ mesons. Given these commonalities, one may wonder  whether the YKG gravitation and, at least the residual strong interactions, are  the two manifestations of a single underlying concept, the two sides of the same coin.

\section*{Acknowledgments}
We thank M. Vahidinia and V. Taghiloo for their participation in formulating the physics of the model elaborated in  Sec.\eqref{def.prob}. We also thank H. Firouzjahi for his insightful comments on the astrophysical aspects of the paper and F. Arash for  discussions on elemental aspects of nucleons mass and their internal structure.

%\section*{References}
%\bibliography{BibliographyFile.bib}

%\bibliography{BibliographyFile.bib}

\begin{thebibliography}{}
%\cite{Sobouti:2023kkb}
\bibitem{Sobouti:2023kkb}
Y.~Sobouti,
``Three arguable concepts: point particle singularity, asymmetric action of EM on quantum wave functions, and the left out restricted Lorentz gauge from U(1),''
Quant. Stud. Math. Found. \textbf{10}, no.2, 223-236 (2023)
doi:10.1007/s40509-022-00290-0
%1 citations counted in INSPIRE as of 24 May 2024

%\cite{Ruffini:1969qy}
\bibitem{Ruffini:1969qy}
R.~Ruffini and S.~Bonazzola,
``Systems of selfgravitating particles in general relativity and the concept of an equation of state,''
Phys. Rev. \textbf{187}, 1767-1783 (1969)
doi:10.1103/PhysRev.187.1767
%967 citations counted in INSPIRE as of 24 May 2024

%\cite{Buchdahl:1959zz}
\bibitem{Buchdahl:1959zz}
H.~A.~Buchdahl,
``General Relativistic Fluid Spheres,''
Phys. Rev. \textbf{116}, 1027 (1959)
doi:10.1103/PhysRev.116.1027
%1023 citations counted in INSPIRE as of 24 May 2024

%\cite{Virbhadra:1997ie}
\bibitem{Virbhadra:1997ie}
K.~S.~Virbhadra,
``Janis-Newman-Winicour and Wyman solutions are the same,''
Int. J. Mod. Phys. A \textbf{12}, 4831-4836 (1997)
doi:10.1142/S0217751X97002577
[arXiv:gr-qc/9701021 [gr-qc]].
%176 citations counted in INSPIRE as of 24 May 2024

%\cite{Moffat:2005si}
\bibitem{Moffat:2005si}
J.~W.~Moffat,
``Scalar-tensor-vector gravity theory,''
JCAP \textbf{03}, 004 (2006)
doi:10.1088/1475-7516/2006/03/004
[arXiv:gr-qc/0506021 [gr-qc]].
%508 citations counted in INSPIRE as of 24 May 2024

%\cite{Moffat:2014pia}
\bibitem{Moffat:2014pia}
J.~W.~Moffat and V.~T.~Toth,
``Rotational velocity curves in the Milky Way as a test of modified gravity,''
Phys. Rev. D \textbf{91}, no.4, 043004 (2015)
doi:10.1103/PhysRevD.91.043004
[arXiv:1411.6701 [astro-ph.GA]].
 %80 citations counted in INSPIRE as of 24 May 2024







%\cite{Jetzer:1992tog}
\bibitem{Jetzer:1992tog}
P.~Jetzer, P.~Liljenberg and B.~S.~Skagerstam,
``Charged boson stars and vacuum instabilities,''
Astropart. Phys. \textbf{1}, 429-448 (1993)
doi:10.1016/0927-6505(93)90008-2
[arXiv:astro-ph/9305014 [astro-ph]].
%26 citations counted in INSPIRE as of 24 May 2024


%\cite{Bezares:2022obu}
\bibitem{Bezares:2022obu}
M.~Bezares, M.~Bo\v{s}kovi\'c, S.~Liebling, C.~Palenzuela, P.~Pani and E.~Barausse,
``Gravitational waves and kicks from the merger of unequal mass, highly compact boson stars,''
Phys. Rev. D \textbf{105}, no.6, 064067 (2022)
doi:10.1103/PhysRevD.105.064067
[arXiv:2201.06113 [gr-qc]].
%49 citations counted in INSPIRE as of 24 May 2024


%\cite{Liddle:1992fmk}
\bibitem{Liddle:1992fmk}
A.~R.~Liddle and M.~S.~Madsen,
``The Structure and formation of boson stars,''
Int. J. Mod. Phys. D \textbf{1}, 101-144 (1992)
doi:10.1142/S0218271892000057
%169 citations counted in INSPIRE as of 24 May 2024





%\cite{Sakharov:1966aja}
\bibitem{Sakharov:1966aja}
 {A.~D.~Sakharov,
``The initial stage of an expanding Universe and the appearance of a nonuniform distribution of matter,''
Sov. Phys. JETP \textbf{22}, 241 (1966)}
%273 citations counted in INSPIRE as of 07 Nov 2024

\bibitem{Gliner1966cvz}
 { E.B. Gliner, `` Algebraic properties of the energy-momentum tensor and vacuum-like states o+ matter''.
  Soviet Journal of Experimental and Theoretical Physics, \textbf{22}, p.378 (1966).}

%\cite{Lan:2023cvz}
\bibitem{Lan:2023cvz}
C.~Lan, H.~Yang, Y.~Guo and Y.~G.~Miao,
``Regular Black Holes: A Short Topic Review,''
Int. J. Theor. Phys. \textbf{62}, no.9, 202 (2023)
doi:10.1007/s10773-023-05454-1
[arXiv:2303.11696 [gr-qc]].
%70 citations counted in INSPIRE as of 07 Nov 2024



%\cite{Georgi:2000vve}
\bibitem{Georgi:2000vve}
H.~Georgi,
``Lie Algebras In Particle Physics : from Isospin To Unified Theories,''
Taylor \& Francis, 2000,
ISBN 978-0-429-96776-4, 978-0-367-09172-9, 978-0-429-49921-0, 978-0-7382-0233-4
doi:10.1201/9780429499210
%27 citations counted in INSPIRE as of 24 May 2024





%\cite{Kerr:2023rpn}
\bibitem{Kerr:2023rpn}
 {R.~P.~Kerr,
``Do Black Holes have Singularities?,''
[arXiv:2312.00841 [gr-qc]].}
%17 citations counted in INSPIRE as of 07 Nov 2024

%\cite{Penrose:1964wq}
\bibitem{Penrose:1964wq}
 {R.~Penrose,
``Gravitational collapse and space-time singularities,''
Phys. Rev. Lett. \textbf{14}, 57-59 (1965)
doi:10.1103/PhysRevLett.14.57}
%1956 citations counted in INSPIRE as of 07 Nov 2024


%\cite{Hawking:1971vc}
\bibitem{Hawking:1971vc}
 {S.~W.~Hawking,
``Black holes in general relativity,''
Commun. Math. Phys. \textbf{25}, 152-166 (1972)
doi:10.1007/BF01877517}
%1280 citations counted in INSPIRE as of 07 Nov 2024




%\cite{Gell-Mann:1968hlm}
%\bibitem{Gell-Mann:1968hlm}
%M.~Gell-Mann, R.~J.~Oakes and B.~Renner,
%``Behavior of current divergences under SU(3) x SU(3),''
%Phys. Rev. \textbf{175}, 2195-2199 (1968)
%doi:10.1103/PhysRev.175.2195
%2262 citations counted in INSPIRE as of 10 Jun 2024




\bibitem{Davis_2019}
B. L. Davis, A. W. Graham, and E. Cameron,
``Black Hole Mass Scaling Relations for Spiral Galaxies. I. MBH–M*,sph ''
 The Astrophysical Journal,\textbf{ 873}(1):85, (2019).

%\cite{Tully:1977fu}
\bibitem{Tully:1977fu}
R.~B.~Tully and J.~R.~Fisher,
``A New method of determining distances to galaxies,''
Astron. Astrophys. \textbf{54}, 661-673 (1977)
%1273 citations counted in INSPIRE as of 24 May 2024





%\cite{Sobouti:2008xz}
\bibitem{Sobouti:2008xz}
Y.~Sobouti,
``Dark companion of baryonic matter,''
[arXiv:0810.2198 [gr-qc]].
%12 citations counted in INSPIRE as of 24 May 2024

\bibitem{shostak1973aperture}
G. S. Shostak,
``Aperture Synthesis Study of Neutral Hydrogen in NGC 2403 and NGC 4236. II. Discussion.,''
Astronomy and Astrophysics,  \textbf{24},  411, (1973).


%\cite{Bosma:1981zz}
\bibitem{Bosma:1981zz}
A.~Bosma,
``21-cm line studies of spiral galaxies. 2. The distribution and kinematics of neutral hydrogen in spiral galaxies of various morphological types.,''
Astron. J. \textbf{86}, 1825 (1981)
%doi:10.1086/113063
%333 citations counted in INSPIRE as of 24 May 2024


%\cite{Begeman:1989kf}
\bibitem{Begeman:1989kf}
K.~G.~Begeman,
``H I rotation curves of spiral galaxies. I - NGC 3198,''
Astron. Astrophys. \textbf{223}, 47-60 (1989)
%262 citations counted in INSPIRE as of 24 May 2024


%\cite{Begeman:1991iy}
\bibitem{Begeman:1991iy}
K.~G.~Begeman, A.~H.~Broeils and R.~H.~Sanders,
``Extended rotation curves of spiral galaxies: Dark haloes and modified dynamics,''
Mon. Not. Roy. Astron. Soc. \textbf{249}, 523 (1991)
doi:10.1093/mnras/249.3.523
%1008 citations counted in INSPIRE as of 24 May 2024


%\cite{Sanders:1998gr}
\bibitem{Sanders:1998gr}
R.~H.~Sanders and M.~A.~W.~Verheijen,
``Rotation curves of uma galaxies in the context of modified newtonian dynamics,''
Astrophys. J. \textbf{503}, 97 (1998)
doi:10.1086/305986
[arXiv:astro-ph/9802240 [astro-ph]].
%136 citations counted in INSPIRE as of 24 May 2024


%\cite{Sanders:2002pf}
\bibitem{Sanders:2002pf}
R.~H.~Sanders and S.~S.~McGaugh,
``Modified Newtonian dynamics as an alternative to dark matter,''
Ann. Rev. Astron. Astrophys. \textbf{40}, 263-317 (2002)
doi:10.1146/annurev.astro.40.060401.093923
[arXiv:astro-ph/0204521 [astro-ph]].
%649 citations counted in INSPIRE as of 24 May 2024




%\cite{Persic:1995tc}
\bibitem{Persic:1995tc}
M.~Persic and P.~Salucci,
``Rotation curves of 967 spiral galaxies,''
Astrophys. J. Suppl. \textbf{99}, 501 (1995)
doi:10.1086/192195
[arXiv:astro-ph/9502091 [astro-ph]].
%112 citations counted in INSPIRE as of 24 May 2024

%\cite{Persic:1995ru}
\bibitem{Persic:1995ru}
M.~Persic, P.~Salucci and F.~Stel,
``The Universal rotation curve of spiral galaxies: 1. The Dark matter connection,''
Mon. Not. Roy. Astron. Soc. \textbf{281}, 27 (1996)
doi:10.1093/mnras/278.1.27
[arXiv:astro-ph/9506004 [astro-ph]].
%956 citations counted in INSPIRE as of 24 May 2024

%\cite{Sobouti:2006rd}
\bibitem{Sobouti:2006rd}
Y.~Sobouti,
``An f(r) gravitation instead of dark matter,''
Astron. Astrophys. \textbf{464}, 921 (2007)
[erratum: Astron. Astrophys. \textbf{472}, 833 (2007)]
doi:10.1051/0004-6361:20077452
[arXiv:0704.3345 [astro-ph]].
%116 citations counted in INSPIRE as of 24 May 2024

  \bibitem{Carroll-17} B.W. Carroll, D.A. Ostlie , ``An introduction to modern astrophysics '', Cambridge University Press, (2017).

\bibitem{Rholf} T.L. Wilson, K. Rohlfs, S. Hüttemeister, ``Tools of Radio Astronomy $($Astronomy and Astrophysics Library$)$'', Springer; 6th ed. (2014).

\bibitem{Kaufmann} R. A. Freedman, W.J. Kaufmann, ``Universe'',  WH Freeman \& Co; 8th edition (2007).




%\cite{Jovanovic:2023tcc}
\bibitem{Jovanovic:2023tcc}
 {P.~Jovanovi\'c, V.~B.~Jovanovi\'c, D.~Borka and A.~F.~Zakharov,
%``Improvement of graviton mass constraints using GRAVITY\textquoteright{}s detection of Schwarzschild precession in the orbit of S2 star around the Galactic Center,''
Phys. Rev. D \textbf{109}, no.6, 064046 (2024)
doi:10.1103/PhysRevD.109.064046
[arXiv:2305.13448 [astro-ph.GA]].}
%7 citations counted in INSPIRE as of 26 Oct 2024

 %\cite{Zakharov:2016lzv}
\bibitem{Zakharov:2016lzv}
 {A.~F.~Zakharov, P.~Jovanovic, D.~Borka and V.~B.~Jovanovic,
``Constraining the range of Yukawa gravity interaction from S2 star orbits II: Bounds on graviton mass,''
JCAP \textbf{05}, 045 (2016)
doi:10.1088/1475-7516/2016/05/045
[arXiv:1605.00913 [gr-qc]].}
%60 citations counted in INSPIRE as of 07 Nov 2024





%\cite{Isham:1971gm}
\bibitem{Isham:1971gm}
C.~J.~Isham, A.~Salam and J.~A.~Strathdee,
``F-dominance of gravity,''
Phys. Rev. D \textbf{3}, 867-873 (1971)
doi:10.1103/PhysRevD.3.867
%321 citations counted in INSPIRE as of 24 May 2024

%\cite{Deser:1981wh}
\bibitem{Deser:1981wh}
S.~Deser, R.~Jackiw and S.~Templeton,
``Topologically Massive Gauge Theories,''
Annals Phys. \textbf{140}, 372-411 (1982)
[erratum: Annals Phys. \textbf{185}, 406 (1988)]
doi:10.1016/0003-4916(82)90164-6
%2995 citations counted in INSPIRE as of 24 May 2024

%\cite{deRham:2010kj}
\bibitem{deRham:2010kj}
C.~de Rham, G.~Gabadadze and A.~J.~Tolley,
``Resummation of Massive Gravity,''
Phys. Rev. Lett. \textbf{106}, 231101 (2011)
doi:10.1103/PhysRevLett.106.231101
[arXiv:1011.1232 [hep-th]].
%1719 citations counted in INSPIRE as of 24 May 2024

%\cite{deRham:2014zqa}
\bibitem{deRham:2014zqa}
C.~de Rham,
``Massive Gravity,''
Living Rev. Rel. \textbf{17}, 7 (2014)
doi:10.12942/lrr-2014-7
[arXiv:1401.4173 [hep-th]].
%997 citations counted in INSPIRE as of 24 May 2024

%\cite{Arkani-Hamed:2002bjr}
\bibitem{Arkani-Hamed:2002bjr}
N.~Arkani-Hamed, H.~Georgi and M.~D.~Schwartz,
``Effective field theory for massive gravitons and gravity in theory space,''
Annals Phys. \textbf{305}, 96-118 (2003)
doi:10.1016/S0003-4916(03)00068-X
[arXiv:hep-th/0210184 [hep-th]].
%822 citations counted in INSPIRE as of 24 May 2024

%\cite{Desai:2018swo}
\bibitem{Desai:2018swo}
S.~Desai,
``Limit on graviton mass from galaxy cluster Abell 1689,''
Phys. Lett. B \textbf{778}, 325-331 (2018)
doi:10.1016/j.physletb.2018.01.052
[arXiv:1708.06502 [astro-ph.CO]].
%44 citations counted in INSPIRE as of 24 May 2024


%\cite{Abell:1989mu}
\bibitem{Abell:1989mu}
G.~O.~Abell, H.~G.~Corwin, Jr. and R.~P.~Olowin,
``A Catalog of rich clusters of galaxies,''
Astrophys. J. Suppl. \textbf{70}, 1 (1989)
doi:10.1086/191333
%934 citations counted in INSPIRE as of 24 May 2024
%\cite{Kerr:2023rpn}


%\cite{DAgostino:2024ojs}
\bibitem{DAgostino:2024ojs}
R.~D'Agostino, K.~Jusufi and S.~Capozziello,
``Testing Yukawa cosmology at the Milky Way and M31 galactic scales,''
Eur. Phys. J. C \textbf{84}, no.4, 386 (2024)
doi:10.1140/epjc/s10052-024-12741-6
[arXiv:2404.01846 [astro-ph.CO]].
%4 citations counted in INSPIRE as of 24 May 2024





%\cite{Freese:2008cz}
\bibitem{Freese:2008cz}
K.~Freese,
``Review of Observational Evidence for Dark Matter in the Universe and in upcoming searches for Dark Stars,''
EAS Publ. Ser. \textbf{36}, 113-126 (2009)
doi:10.1051/eas/0936016
[arXiv:0812.4005 [astro-ph]].
%123 citations counted in INSPIRE as of 24 May 2024


\end{thebibliography}
\newpage

\end{document}